\documentclass[runningheads]{llncs}

\usepackage[T1]{fontenc}

\usepackage{float}
\usepackage{graphicx}
\usepackage{url}
\usepackage{hyperref}
\usepackage[hyphenbreaks]{breakurl}
\usepackage{xspace}

\usepackage{wrapfig}

\usepackage{counters}
\renewcommand{\reffig}[1]{Fig.~\ref{fig:#1}}
\renewcommand{\refeqn}[1]{(\ref{eqn:#1})}
\renewcommand{\refrule}[1]{Rule~(\ref{rule:#1})}
\usepackage{xcolor}
\usepackage{amsmath}
\usepackage{relsize}
\usepackage{soundness}
\usepackage{rcoms}
\usepackage{wmm}

\usepackage{isabelle}
\usepackage{isabellesym}

\usepackage{ifthen}

\usepackage{oz}
\renewcommand{\spot}{\mathbin{.}}
\newcommand{\cons}{\mathbin{\#}}

\newenvironment{clhproof}[1][]{
	{\bf Proof.}
}{
	$\qed$
}

\renewcommand{\ttdef}{::=}

\newcommand{\Blue}[1]{{\color{blue}#1}}
\newcommand{\procname}[1]{\mathsf{#1}}
\newcommand{\specvar}[1]{\Blue{#1}}
\newcommand{\clhconst}[1]{\mathsf{#1}}

\newcommand{\acquireproc}{\procname{acquire}}
\newcommand{\releaseproc}{\procname{release}}
\newcommand{\lock}{\specvar{lock}}

\newcommand{\acquirei}{\acquireproc_\tidi()}
\newcommand{\releasei}{\releaseproc_\tidi()}

\newcommand{\free}{\clhconst{free}}
\newcommand{\held}[1]{\clhconst{held}(#1)}
\newcommand{\heldi}{\held{\tidi}}

\newcommand{\tid}[1]{\mathsf{#1}}
\newcommand{\tidi}{\tid{i}}
\newcommand{\tidj}{\tid{j}}

\def\codeindent{~~}

\newcommand{\elist}{[\,]}

\newcommand{\clhsel}{CLH-seL4\xspace}

\newcommand{\Thread}{\mathcal{T}}

\newcommand{\kw}[1]{{\color{purple}\mathsf{#1}}}
\newcommand{\Array}{\mathop{\kw{array}}}

\renewcommand{\Do}{\mathrel{\kw{do}}}
\renewcommand{\Else}{\mathop{\kw{else}}}

\newcommand{\guar}{\kw{guar}}

\renewcommand{\Init}{\mathop{\kw{initially}}}
\newcommand{\Invariant}{\mathop{\kw{invariant}}}

\newcommand{\Of}{\mathbin{\kw{of}}}
\newcommand{\rely}{\kw{rely}}
\renewcommand{\Then}{\mathop{\kw{then}}}
\newcommand{\Type}{\mathop{\kw{type}}}
\newcommand{\Var}{\mathop{\kw{var}}}

\renewcommand{\While}{\kw{while}~}
\renewcommand{\Await}{\kw{await}~}

\newcommand{\pmid}{p_{mid}}
\newcommand{\labelinfrule}[1]{\label{infrule:#1}}
\newcommand{\refinfrule}[1]{Rule~(\ref{infrule:#1})}
\newcommand{\rginfrule}[3]{
\begin{minipage}[b]{#1\textwidth}
\begin{equation}
	\labelinfrule{#2}
	\hskip -1mm
	#3
\end{equation}
\end{minipage}}

\newcommand{\swapproc}{\texttt{swap}}
\newcommand{\swap}[2]{\swapproc(#1,#2)}
\newcommand{\ID}[1]{\mathsf{id}(#1)}

\newcommand{\previ}{\prev_\tidi}

\newcommand{\reservedi}{\reserved[\tidi]}
\newcommand{\awaitgranted}{
	\Await \getstatus{\prev} = \Granted 
}

\newcommand{\curnodei}{\procs[\tidi].\curnode}
\newcommand{\nextnodei}{\procs[\tidi].\nextnode}

\newcommand{\auxil}[1]{\red{#1}}

\newcommand{\getstatus}[1]{#1\rightarrow\status}

\newcommand{\llangle}{\langle\!\langle}
\newcommand{\rrangle}{\rangle\!\rangle}

\renewcommand{\last}{\clhconst{last}}
\newcommand{\butlast}{\clhconst{butlast}}
\newcommand{\fmap}[2]{#1\llangle#2\rrangle}
\newcommand{\distinct}[1]{\clhconst{distinct}(#1)}
\newcommand{\distinctq}{\distinct{\queue}}

\newcommand{\Seq}{\mathop{;}}

\newcommand{\concrete}[1]{\mathtt{#1}}
\newcommand{\typename}[1]{\mathtt{#1}}

\newcommand{\injective}[1]{injective(#1)}

\newcommand{\status}{\concrete{status}}
\renewcommand{\tail}{\concrete{tail}}
\newcommand{\curnode}{\concrete{node}}
\newcommand{\nextnode}{\concrete{tmp}}
\newcommand{\auxhead}{\auxil{\eta}}
\newcommand{\prev}{\concrete{prev}}
\renewcommand{\p}{\concrete{p}}
\newcommand{\reserved}{\auxil{reserved}}

\newcommand{\StatusVal}{\typename{Status}}
\newcommand{\Node}{\typename{Node}}

\newcommand{\Granted}{\clhconst{Granted}}
\newcommand{\Pending}{\clhconst{Pending}}

\newcommand{\numThreads}{N}

\renewcommand{\self}{\tidi}

\newcommand{\nothing}{\mathop{\kw{skip}}}

\newcommand{\Assert}[1]{~~~\inlineAssert{#1}}
\newcommand{\inlineAssert}[1]{~\green{\{#1\}}~}

\newcommand{\contract}[1]{\kw{contract}_{#1}}
\newcommand{\contracti}{\contract{\tidi}}
\newcommand{\contractj}{\contract{\tidj}}
\newcommand{\cinv}{\kw{cinv}}

\newcommand{\indexin}[2]{#1!#2}

\newcommand{\stable}[2]{#1~\mathsf{stable}~\mathsf{under}~#2}
\newcommand{\stablep}[1]{\stable{p}{#1}}

\newcommand{\stablepr}{\stablep{r}}

\newcommand{\dti}{I}
\newcommand{\dtir}{\makerel{\dti}}

\def\invsep{\slash\!\!\slash}
\newcommand{\quintI}[6]{\quint{#1}{#2}{#3\invsep #6}{#4}{#5}}
\newcommand{\quinti}[5]{\quintI{#1}{#2}{#3}{#4}{#5}{\dti}}
\newcommand{\quintiprgqc}{\quinti{p}{r}{c}{g}{q}}

\begin{document}

\title{Practical Rely/Guarantee Verification of an Efficient Lock for seL4 on Multicore Architectures}
\titlerunning{Rely-guarantee verification of locks for multicore architectures}

\author{Robert J. Colvin \inst{1,2} \and
Ian J. Hayes\inst{2} \and
Scott Heiner\inst{2} \and \\
Peter H\"ofner\inst{3} \and
Larissa Meinicke\inst{2} \and
Roger C. Su\inst{3}
}

\authorrunning{R. Colvin et al.}

\institute{
Defence Science and Technology Group, Australia
\and
University of Queensland, Brisbane, Australia \\
\email{\{r.colvin, ian.hayes, s.heiner, l.meinicke\}@uq.edu.au}
\and
Australian National University, Canberra, Australia \\
\email{\{peter.hoefner, roger.su\}@anu.edu.au}
}

\maketitle

\begin{abstract}

Developers of low-level systems code providing core functionality for operating systems and kernels must address 
hardware-level features of modern multicore architectures.
A particular feature is pipelined ``out-of-order execution'' of the code as written,
the effects of which are typically summarised as a ``weak memory model'' --
a term which includes
further complicating factors that may be introduced by compiler optimisations.
In many cases, the nondeterminism inherent in weak memory models can be expressed as micro-parallelism, \ie
parallelism within threads and not just between them.
Fortunately Jones' \emph{rely/guarantee} reasoning provides a compositional method for shared-variable concurrency, 
whether that be in terms of communication between top-level threads or micro-parallelism within threads.
In this paper we provide an in-depth verification of the lock algorithm used in the seL4 microkernel,
using rely/guarantee to handle both interthread communication as well as micro-parallelism introduced
by weak memory models.


\end{abstract}

\setcounter{footnote}{0}

\section{Introduction}
The \emph{CLH lock}
\cite{craig-93,MLH-94-tech,MLH-94}
provides a space- and time-efficient method for ensuring mutual exclusion,
and furthermore ensures that any thread attempting to acquire the lock will eventually succeed
under reasonable fairness assumptions.  CLH's efficiency arises partly from its
utilisation of the hardware-level cache system to minimise contention in main memory.
It is used in the seL4 microkernel~\cite{seL4-Commun-ACM,CLH-seL4} 
to handle interrupts.

In this paper, we
formally verify the functional correctness of the CLH lock,
and lay the foundation for proving progress properties. 
Additionally, we take into account the effects of
\emph{weak memory models} of possible target architectures for the deployment of seL4.
We use rely/guarantee reasoning for this task, and find it especially suited for hardware-level algorithms which involve weak memory effects.
We use the proof assistant Isabelle/HOL~\cite{IsabelleHOL,Paulson:94}, and build on existing library features (including a rely/guarantee encoding by Prensa Nieto~\cite{RGinIsabelle}) to automate some parts of the proof.

The paper is structured as follows:
\refsect{rely-guarantee} summarises Jones' rely/guarantee reasoning, with some specialisations for our context.
\refsect{lock-specification} introduces the specification of a generic lock, and its specialisation to
a queued lock.
\refsect{CLH-seL4-proof} proves functional correctness of the CLH lock currently used in the seL4 microkernel.
with respect to the specification of a queued lock.
\refsect{wmms}
analyses weak memory effects on the CLH code, 
identifying the micro-parallelism that may be introduced during execution on multicore architectures,
and whether or not that micro-parallelism may be retained or must be eliminated.
\refsect{related-work} summarises related work.



\newcommand{\updatepred}[3]{#1_{\langle #2 \leftarrow #3\rangle}}
\newcommand{\updaterel}[2]{#1 \leftarrow #2}
\newcommand{\poststate}[1]{\mathsf{post}(#1)}
\newcommand{\genpar}[2]{\parallel_{#1}\!.~{#2}}

\section{Rely/Guarantee Reasoning}
\labelsect{rely-guarantee}

Developed by Jones in the 1980s~\cite{Jones-RG0,Jones-RG1,Jones-RG2}, rely/guarantee reasoning provides a compositional framework for the verification of programs with
shared-variable concurrency. The key idea is to extend pre/postcondition reasoning of classic Floyd-Hoare logic~\cite{Floyd_67,Hoare_69}
by capturing the effect of the environment of a thread in a
\emph{rely} relation on states; additionally, a thread
summarises its own effects on its environment via a \emph{guarantee} relation on states. 

In the context of a
lock, a thread $\tidi$ \emph{relies} on the fact that if $\tidi$ currently holds the lock
then some resource $x$ cannot be 
modified by the
environment. 
This is written as $\lock = \heldi \imp x' = x$, a rely \emph{relation}
where $\lock$ and $x$ are shared variables, 
with primed instances ($x'$) referring to their values
in the \emph{post-state} (as opposed unprimed instances ($x$) in the \emph{pre-state}).
The corresponding \emph{guarantee} relation is
$\lock \neq \heldi \imp x' = x$;
that is,
thread $\tidi$ guarantees not to change $x$ if it does not hold the lock.
In this work, pre/postconditions are
single-state predicates (where a state is a mapping from variables to values),
and relies and guarantees are binary relations on states (with primed as well as unprimed variables).
We allow the standard logical connectives to apply to either predicates or relations under their usual interpretations. 

A rely/guarantee quintuple $\quintprgqc$
states that the program $c$ achieves the postcondition $q$ and guarantees each step satisfies $g$, provided that it starts in a state satisfying $p$
and each step of the environment satisfies the rely $r$.
Inference using rely/guarantee takes place via the application of specialised syntax-driven rules.
The rules we apply are adaptations of some of the rules provided by Coleman and Jones~\cite{SoundnessRG} 
and those encoded in Isabelle by Prensa Nieto~\cite{RGinIsabelle}.

\subsection{Inference rules}

\begin{figure}

\rginfrule{0.33}{asgn}{
	\Rule{
		\stable{p,q}{r}
		\\
		\updatepred{p}{x}{e} \imp q
		\\
		p \dres (x' = e) \imp g
	}{
		\quintpr{x \asgn e}{g \lor \id}{q}
	}
}
\quad 
\rginfrule{0.29}{seq}{
	\Rule{
		\quint{p}{r}{c_1}{g}{\pmid}
		\\
		\quint{\pmid}{r}{c_2}{g}{q}
	}{
		\quint{p}{r}{c_1 \Seq c_2}{g}{q}
	}
}
\quad
\rginfrule{0.29}{conseq}{
	\Rule{
		\quint{p_0}{r_0}{c}{g_0}{q_0}
		\\
		p \imp p_0
		\quad
		r \imp r_0
		\\
		g_0 \imp g
		\quad
		q_0 \imp q
	}{
		\quint{p}{r}{c}{g}{q}
	}
}

\rginfrule{0.43}{par-u}{
	\Rule{
		\quint{p}{r \lor g_2}{c_1}{g_1}{q_1}
		\\
		\quint{p}{r \lor g_1}{c_2}{g_2}{q_2}
	}{
		\quint{p}{r}{c_1 \pl c_2}{g_1 \lor g_2}{q_1 \land q_2}
	}
}
\hskip -1mm
\rginfrule{0.56}{par-int}{
	\Rule{
		\quint{p_1}{r_1}{c_1}{g_1}{q_1}
		\\
		\quint{p_2}{r_2}{c_2}{g_2}{q_2}
		\\
		g_1 \imp r_2 \land g_2 \imp r_1
	}{
		\quint{p_1 \land p_2}{r_1 \land r_2}{c_1 \pl c_2}{g_1 \lor g_2}{q_1 \land q_2}
	}
}

\rginfrule{0.47}{par-gen}{
	\Rule{
		\forall i \spot \quint{p_i}{r_i}{c_i}{g_i}{q_i}
		\\
		\forall i \spot \forall j \neq i \spot g_i \imp r_j
	}{
		\quint
			{\bigwedge_i p_i}
			{\bigwedge_i r_i}
			{\genpar{i}{c_i}}
			{\bigvee_i g_i}
			{\bigwedge_i q_i}
	}
}
\qquad \qquad
\rginfrule{0.40}{spin-loop}{
	\Rule{
		\stable{p,q}{r}
		\\
		\id \imp g
		\qquad
		p \land \neg b \imp q
	}{
		\quintprgq{\WHb{b}~\nothing}
	}
}

\caption{Rely/guarantee inference rules adapted from 
\cite{SoundnessRG,RGinIsabelle}.
}
\labelfig{infrules}

\end{figure}

The inference rules we apply in this paper are given in \reffig{infrules}.
\refinfrule{asgn}
states that an atomic assignment $x \asgn e$ establishes $q$ and guarantees $g \lor \id$
under precondition $p$ and rely $r$,
provided that neither $p$ nor $q$ is affected by the environment,
and executing $x \asgn e$ from a state satisfying $p$ 
both establishes $q$ in the post state and implies the guarantee $g$.
As an example, we may immediately derive
$
	\quint
		{\True}
		{x' = x}
		{x \asgn 5}
		{\updaterel{x}{5} \lor \id}
		{x = 5}
$.
The inclusion of the identity relation $\id$ to the guarantee is a technical requirement that accounts
for the fact that the assignment statement inflicts no interference on the rest of the system prior to executing --
between steps of $c$ in `$c \pl x \asgn 1$'
either the assignment occurs or nothing ($\id$) occurs.
We let
$\updatepred{p}{x}{e}$ be the predicate obtained by updating each state in $p$ so that $x$ takes the value of $e$ in the pre-state;
$p \dres r$ be the restriction of the relation $r$ to just the pre-states satisfying predicate $p$;
and
$\stablepr$ states that $p$ is maintained by $r$.

\refinfrule{seq} composes the properties of two commands in sequential composition provided they agree on a \emph{mid-state},
written $\pmid$ in the rule.  The relies and guarantees are identical across both components.

\refinfrule{conseq} is a general rule allowing the weakening of preconditions and relies, and strengthening of postconditions and guarantees.
We tend to leave the application of this rule implicit where the implications are clear from the context.

The significance of rely/guarantee lies in its handling of parallel composition.
There are two compositional rules typically found in the literature that apply to two parallel processes, $c_1 \pl c_2$.
\refinfrule{par-u} assumes some external environment ``inflicting'' $r$ on the local system (in this case, the local
system is $c_1 \pl c_2$).
The compositionality of the rule requires that the additional interference $g_2$ inflicted on $c_1$ by $c_2$ must be 
handled (by $c_1$),
and vice-versa.  
\refinfrule{par-int} instead insists that the environment must already respect the relies of both threads $c_1$ and $c_2$,
and each thread must also respect each other's relies.
We found that each has its application; in our case, micro-parallelism within a single thread (as addressed in \refsect{wmms}) 
is best handled by \refinfrule{par-u},
while at the system level, where a set of threads interact via a shared contract, \refinfrule{par-int} is more suitable.
This is given by the generalisation of \refinfrule{par-int} to
an arbitrary number of threads, written
$
\genpar{i}{c_i}
$, as shown in \refinfrule{par-gen}.

A common structure in lock implementations are \emph{spin loops}, 
i.e., loops with an empty body, $\WHb{b}\nothing$.
We often abbreviate spin loops to $\Await \neg b$.
\refinfrule{spin-loop} states that
provided the precondition $p$ and postcondition $q$ are stable under $r$,
and that $p$ and the negation of the guard imply $q$,
one can assume that $q$ holds if the loop terminates.
A spin loop changes nothing, and hence guarantees the identity relation, $\id$.

\subsection{Invariants}

\renewcommand{\dtir}{\dti \imp \dti'}

We define the following shorthand for a (single-state) \emph{data-type invariant} $\dti$
(or any invariant), in the spirit of Morgan's refinement calculus \cite{Morgan:94}.
\begin{eqnarray}
	\quintiprgqc
	\asdef
	\quint{p \land \dti}{r \land (\dtir)}{c}{g \land (\dtir)}{q \land \dti}
	\labeleqn{invariant-quintuple}
\end{eqnarray}
Definition \refeqn{invariant-quintuple} factors out the invariant, making specification more concise.
It states that if $\dti$ holds in the pre-state, and the environment can be relied upon to maintain the invariant,
then $c$ guarantees to maintain the invariant and establish it in the post-state
(the relation $\dtir$ requires that $\dti$ holds in the post-state provided it holds in the pre-state).

The use of an invariant helps to avoid the use of code-specific program-counter-style variables, particularly when used on auxiliary concepts (essentially forming the ``specification'' of a data type).  Program-counters are preferably avoided, as 
they are brittle with respect to
the structure of the code, and become cumbersome with complex programs, such as those with nested parallelisms.



\newcommand{\idlock}[1]{\lock = \held{#1} \implies \lock' = \held{#1}}
\newcommand{\idlocki}{\idlock{\tidi}}
\newcommand{\idlockj}{\idlock{\tidj }}

\newcommand{\stayinq}[1]{#1 \in \queue \iff #1 \in \queue'}
\newcommand{\stayinqi}{\stayinq{\tidi}}
\newcommand{\stayinqj}{\stayinq{\tidj}}

\newcommand{\queue}{\specvar{q}}
\renewcommand{\self}{\mathsf{\tidi}}

\newcommand{\hd}[1]{\clhconst{hd}(#1)}
\newcommand{\hdq}{\hd{\queue}}
\newcommand{\tl}[1]{\clhconst{tl}(#1)}
\newcommand{\tlq}{\tl{\queue}}

\newcommand{\LockType}{\typename{LockStatus}}

\section{Specification of Locks in Rely/Guarantee}
\labelsect{lock-specification}

In this section we discuss specifying locks in a rely/guarantee setting.
The auxiliary variable $\lock \in \LockType$ represents the status of the lock,
which is either $\free$ if no thread holds the lock, or $\heldi$ if thread $\tidi$ holds the lock.

\subsection{Properties of a generic lock}
\labelsect{lock-props}

A lock is used by multiple threads (type $\Thread$) to ensure exclusive access to some \emph{critical section}
of code, typically where some shared resource is accessed.
That is, from the viewpoint of each thread $\tidi \in \Thread$, the lock is used in the pattern%
\footnote{Note that $\acquireproc$ and $\releaseproc$ have no explicit parameters,
but in some implementations they may
(usually some component of the lock data structure).
They do have the thread identifier ($\tidi$) as an implicit parameter.}
\begin{equation*}
\acquirei \Seq \emph{critical~section} \Seq \releasei.%
\end{equation*}
The intention is that if modifying $x$ in the critical section,
a thread may rely on the fact that no other thread concurrently modifies $x$,
that is, in the calling code,
$\lock = \heldi \imp x' = x$.  

The operation $\acquirei$ is called to attempt to hold the lock, 
and $\releasei$ to release the lock after the thread has finished accessing the shared resource.
The key properties required are the following.
\begin{enumerate}
\item
A call to $\acquirei$ establishes the postcondition 
$\lock = \heldi$; 
and
\item
Thread $\tidi$ may \emph{rely} on 
$
	\lock = \heldi \imp \lock' = \heldi
$.
\end{enumerate}
The postcondition 
$
\lock = \heldi
$
is \emph{stable} -- maintained by the environment --
if the environment satisfies the rely,
that is,
if $\tidi$ holds the lock then it will never be taken away.
These conditions necessitate the converse guarantee, that thread $\tidi$ will respect the lock
if some other thread $\tidj$ holds it.

\subsection{Queued lock}

When multiple threads call $\acquirei$ concurrently, a potentially desirable property is that the lock is next obtained
by the thread that has been waiting the longest.
To capture this refinement of the lock concept
a generic lock can be modelled via an abstract queue of waiting threads, $\queue$, 
as given in \reffig{queued-lock-spec}.
We let $\tidi \in \queue$ mean that $\tidi$ is an element of list $\queue$, and let $\hdq$ return $\queue$'s first element.
\begin{figure}[t]

\begin{equation}
	\inlineAssert{\tidi \notin \queue}
	\acquirei
	;
	\inlineAssert{\tidi = \hdq}
	\releasei
	\inlineAssert{\tidi \notin \queue}
\end{equation}
\begin{eqnarray}
	\Invariant &\sdef&
	\distinctq 
	\\
	\contracti
	\asdef
	(\stayinqi) \land
	(
	\self \in \queue \imp
	\indexin{\queue'}{\tidi} 
	\leq
	\indexin{\queue}{\tidi} 
	)
	\\
	\rely{\tidi}
	\asdef
	\Invariant \land \contracti
	\\
	\guar{\tidi}
	\asdef
	\Invariant \land (\forall \tidj \neq \tidi \spot \contract{\tidj})
	\\
	\cinv &\sdef&
	\lock = (\If \queue = \elist \Then \free \Else \held{\hdq})
\end{eqnarray}

\caption{Rely/guarantee specification of the queued lock}
\labelfig{queued-lock-spec}
\end{figure}

To emphasise the different parts of a specification, 
we separate out the guarantees in $\guar$ and the relies in $\rely$,
and the pre- and postconditions are written in green braces.
Hence, \reffig{queued-lock-spec} implies a quintuple of the form in \refeqn{invariant-quintuple}, \\
\centerline{
$\quintI{\tidi \notin \queue}{\rely_\tidi}{\acquirei \Seq \releasei}{\guar_\tidi}{\tidi \notin \queue}{\Invariant}$.
}
\\
The intermediate states are added to the structure of the program, in this case, between $\acquirei$ and $\releasei$,
representing predicates that can be used to fulfil the antecedents of \refinfrule{seq} during a proof of the full quintuple.
For instance, the postcondition for $\acquirei$ also forms the precondition for $\releasei$,
which is apparent from the structure of code interspersed with states (similar to an Owicki-Gries-structured proof
\cite{OGinIsabelle,OwickiGries76}).
The data type $\Invariant$, that each thread identifier in $\queue$ can appear at most once, is maintained throughout.

The specification 
introduces a `$\contracti$' for thread $\tidi$,
which is a relation describing thread $\tidi$'s assumption about how and when the state will be modified.
Then each thread guarantees to fulfil the contract for every other thread in the system.
In this case, the contract states that $\tidi$ cannot be entered or removed from the queue by any other thread,
and additionally, that once in the queue it can only ever get closer to the head:
$
	\self \in \queue \imp
	\indexin{\queue'}{\tidi} 
	\leq
	\indexin{\queue}{\tidi} 
$.
We use the notation $\indexin{\queue}{\tidi}$ to denote the index of the element $\tidi$ in the list $\queue$.
This expression is undefined if the list $\queue$ does not contain $\tidi$;
we ensure throughout that such indexing is defined.

The key part establishing that this specification meets that of a lock is the
coupling invariant $\cinv$ which \emph{derives} the lock state from the queue:
if the queue is empty then the lock is $\free$, and otherwise the first thread in the queue holds the lock.
Under this definition, the predicates and relations in \reffig{queued-lock-spec} imply the lock properties given
in \refsect{lock-props}.
We use a similar justification later for showing that CLH implements an (abstract) lock.



\newcommand{\Nodeptr}{*\Node}

\newcommand{\NodeOwner}{\concrete{proc}}
\newcommand{\procs}{\concrete{procs}}

\newcommand{\addr}{\mbox{\tt \&}}

\newcommand{\cstruct}{\mathtt{struct}}

\section{Functional Correctness of CLH-seL4}
\labelsect{CLH-seL4-proof}

The seL4 microkernel was the first operating system to be formally verified for functional correctness \cite{seL4-verification},
and its development continues \cite{seL4-Commun-ACM},
including formal proofs of security criteria \cite{seL4-information-flow}.
The original proof of correctness essentially assumed execution on a single core,
but
recent developments have seen deployment on multicore Arm, and potentially x86 and RISC-V, necessitating the need for managing concurrent processors.
This is addressed via a system-wide lock to handle interrupts, and this lock is currently implemented via a version of CLH~\cite{CLH-seL4}.
CLH is an interesting target for verification via rely/guarantee, because it is efficient (used in practice) and involves
fine-grained concurrency that cannot be managed by programming language abstractions: it operates at the microkernel level and therefore cannot
assume access to library code, \etc

We first present the acquire and release operations in a \Clang-like language corresponding to the implementation currently in the
release version of seL4.
We apply rely/guarantee reasoning to this code, augmented with auxiliary variables capturing key concepts,
and prove that it satisfies the conditions for a concurrent lock.
We assume lines of code are executed in the order specified (\emph{sequential consistency}) --  weak memory
effects are addressed in \refsect{wmms}.

\subsection{CLH-seL4 implementation}

\reffig{CLH-seL4-code} contains code similar to that used in the seL4 microkernel,
with interrupt-handling specifics stripped out (that code is separate to the locking mechanism;
we have also renamed some variables for consistency with other presentations of CLH).
Broadly speaking,
the code maintains a ``queue'' of threads in a similar fashion to a linked list structure through the local $\prev$ variables.
We give a brief description below,
but defer a full explanation of the operation of the code until we augment it with the key auxiliary concepts.
The verification does not need to be aware of cache line sizes, \etc, as that is an
efficiency concern.

\begin{figure}[ht]
\[
	\mathtt{
  \Type \StatusVal = \Granted | \Pending
  	} \qquad \mathtt{
  \Type \Node = \cstruct \{ \status: \StatusVal \}
  	} \\ \mathtt{
  \Type \NodeOwner = \cstruct \{ \curnode: *\Node ; \nextnode: *\Node \}
 	} \\
  \also
 \begin{array}{l}
  \mathtt{
  \Var \procs : \Array \numThreads \Of \NodeOwner \asgn \ldots  ;
  } \qquad \mathtt{
  \Var \tail: *\Node \asgn 
  	\ldots
  } \\ \mathtt{
  	\getstatus{\tail} \asgn \Granted ;
  } 
 \end{array}
 \\
 \also
 \hskip -5mm
 \begin{array}{l}
 	\mathtt{
  \acquireproc(\self : \Thread) = 
  } \\ 
 \begin{array}{@{\codeindent}l}
 \mathtt{
   \getstatus{\curnodei} := \Pending; 
   } \\ \mathtt{
   \Var \prev \asgn \swap{\addr\tail}{\curnodei} ; 
   } \\ \mathtt{
   \nextnodei \asgn \prev ; 
   } \\ \mathtt{
   \awaitgranted \Seq 
   }
 \end{array} 
 \end{array} 
	\begin{array}{l}
 	\mathtt{
  \releaseproc(\self : \Thread) = 
  } \\ 
 \begin{array}{@{\codeindent}l}
 \mathtt{
    \getstatus{\curnodei} \asgn \Granted ; 
	} \\ \mathtt{
	\curnodei \asgn \nextnodei
	}
 \end{array} 
 \end{array}
\]

\caption{Excerpt of the seL4 CLH implementation (focussing on the logic) \cite{CLH-seL4}.}
\labelfig{CLH-seL4-code}
\end{figure}

\paragraph{Types and initialisation.}
The code assumes a fixed number of threads, $\numThreads$,
and each thread $\tidi$ is initially allocated a pointer to its own $\Node$, which can be accessed via $\curnodei$.
The node itself is simply a boolean which indicates whether the next waiting node must continue to wait ($\Pending$) or may
take the lock ($\Granted$).
There is one more $\Node$ than there are threads, with this extra node effectively marking the head of the abstract queue.
The array $\procs$ is initialised so that each thread has a unique node,
and $\tail$ is initialised to the unique unused node
(we omit these straightforward details).
Initially, the status of $\tail$ 
is set to $\Granted$ so that the first node to attempt to acquire the lock may do so.

\paragraph{Acquire.}
To acquire the lock a thread $\tidi$ initialises its node to have the status $\Pending$, indicating to the subsequent thread $\tidj$
that joins the queue that $\tidj$ cannot 
(immediately) take the lock.
An \emph{atomic swap} operation is then called, which reads and returns the current value of the $\tail$ pointer, and updates $\tail$
to the node of the calling thread.
Thus, the $\prev$ local variable records the previous node in the queue,
and this is 
then stored in $\nextnodei$ for access in a call to $\releasei$. This $\prev$ node will eventually be used by $\tidi$ the next time it calls $\acquireproc$.
The thread then spins on the status of the previous node, waiting for it to become $\Granted$, at which point thread $\tidi$ takes the lock
and enters its critical section.
Recall that the $\Await b$ command abbreviates a while loop with an empty body, 
i.e.,
$
	\While \neg b \Do \nothing
$.

\paragraph{Release.}
To release the lock, the thread sets the status of its own node to $\Granted$, signalling to the next waiting thread that it may proceed.
It also now prepares for its next call to $\acquirei$ by copying the node saved in $\nextnodei$ into $\curnodei$.
Note that another thread may concurrently take the lock and proceed to its critical section before thread $\tidi$ executes this second line of $\releasei$.

\paragraph{Efficiency.}
In terms of space efficiency, one $\Node$ is allocated per-processor
initially, plus one additional node required to track the head of an empty queue.  
Threads swap nodes between themselves during execution, but the complete set of nodes never
changes -- there is no need for freeing/allocating memory during execution (as long as the total number of threads that will access the lock is known beforehand).

The second aspect of efficiency involves a specific interaction with the memory (cache) system.
Although the status of a node can be recorded in a single bit, the $\status$ array
is padded out so that each boolean takes up space equal to a single cache line -- this is the physical unit that
processors may read from main memory.  Even if a single byte is requested from main memory, the entire cache line
in which that byte resides is read into the processor's local cache.  The size of a cache line is processor-specific;
64 bytes is a typical size used in contemporary architectures.  
As long as no process in the system changes any part of that 64 bytes, the local
process can read from its copy of that chunk without accessing shared memory, which would otherwise
potentially slow down the entire system.
CLH exploits this so that each processor's spin loop is effectively spinning on its own local cache, completely separately
to every other process in the system, and avoiding repeated requests to main memory to retrieve a value.
The only time that main memory is accessed is precisely when the previous processor in the queue flips the status
from $\Pending$ to $\Granted$, triggering a refresh of that cache line.

\subsection{Auxiliary variables, invariants and intermediate states of CLH-seL4}

\reffig{CLH-seL4-proof} outlines the proof of the CLH-seL4, 
including the invariant, pre/post\-condi\-tion conditions, rely/guarantees, and intermediate states. 
The proof uses the following notation, where
$a \cons t$ represents a non-empty list with first (head) element $a$,
and `$\cat$' is list concatenation.  
\begin{eqnarray*}
	\injective{f} \asdef 
  		\forall i\ j \in \dom f \spot i \neq j \implies f(i) \neq f(j)  
	\labeleqn{defn-injective}
	\\
	\last(t\cat[a]) = a
	\qquad &&\qquad
	\butlast(t\cat[a]) = t
	\\
	\fmap{f}{\elist} = \elist
	&&
		\fmap{f}{a\cons t} = f(a) \cons  (\fmap{f}{t}) 
		\\
	\distinct{t} \asdef length(t) = size(set(t))
\end{eqnarray*}

\paragraph{Types, initialisation, and auxiliaries.}
We augment the code with several auxiliary variables to capture key ideas.
The $\cstruct$s are carried over from \reffig{CLH-seL4-code}.
There is one local variable $\prev$ per thread, which we represent in the usual way as a mapping from thread identifiers to the type of the local --
in this case, $\Nodeptr$.
As such the $\prev$ for thread $\tidi$ is written $\prev_\tidi$, although there is otherwise no distinction between 
a local variable and an array indexed by threads.
Wherever nodes are dereferenced -- to access their $\status$ -- we implicitly access the $heap$, which 
is a mapping from $\Nodeptr$ to $\Node$.

The most complex part of the algorithm is how nodes are shared between threads.
The auxiliary variable $\reserved$, which by convention we write in \red{red}, keeps track of the unique node currently marked by a process for its own use.
This auxiliary concept mostly corresponds to the concrete $\curnodei$, except for a brief period
during $\releaseproc$, where a thread's reserved node may be found in $\nextnodei$.
In a symbiotic way, the auxiliary $\auxhead$ variable records the lone unreserved node in the system,
which is always the previous ($\prev$) node of the thread holding the lock,
or $\tail$ if the queue is empty.

We also maintain an abstract list of the waiting threads in $\queue$.  While this value could be \emph{derived} from the concrete state,
making it explicit in the code also acts as a type of abstract ``program counter'', indicating
which threads are currently attempting to acquire the lock.

\begin{figure}[H]
\[
 \begin{array}{ll}
   \Var \prev : \Thread \fun \Nodeptr; 
   	& (uninitialised~local)
  \\
  \Var \reserved : \Array \Thread \Of \Nodeptr ;
  	& \Init~\lambda \tidi \spot \curnodei \\ 
  \Var \auxhead : \Nodeptr ;
  	& \Init~\tail 
\\
  \Var \queue : \Thread~list ;
  	& \Init~\elist
 \end{array}
\]

\[
\begin{array}{ll}
  \acquirei \sdef 
  \\
 \begin{array}{@{\codeindent}l}
   \Assert{\self \notin \queue \land \curnodei = \reservedi} \\
   \getstatus{\curnodei} := \Pending; \\
   \Assert{\self \notin \queue \land \curnodei = \reservedi \land \getstatus{\curnodei} = \Pending} \\
   \atomic{\prev \asgn \tail ; \tail \asgn \curnodei ; \queue \asgn \queue \cat [\self]} ; \\
   \Assert{\self \in \queue \land \curnodei = \reservedi} \\
   \nextnodei \asgn \prev ; \\
   \Assert{\self \in \queue \land \curnodei = \reservedi \land \nextnodei = \prev} \\
   \awaitgranted \Seq \\
   \Assert{\self \in \queue \land \curnodei = \reservedi \land \nextnodei = \auxhead \land \self = \hdq} \\
 \end{array}
  \also
  \releasei  \sdef
  \\
 \begin{array}{@{\codeindent}l}
   \Assert{\self \in \queue \land \curnodei = \reservedi \land \nextnodei = \auxhead \land \self = \hdq} \\
    \atomic{
		\getstatus{\curnodei} \asgn \Granted \Seq 
		\auxhead \rightleftharpoons \reservedi \Seq
		\queue \asgn \tlq 
	} \Seq \\
	\Assert{\self \notin \queue \land \nextnodei = \reservedi} \\
	\curnodei \asgn \nextnodei \\
	\Assert{\self \notin \queue \land \curnodei = \reservedi} 
 \end{array}
 \\
 \end{array}
\]

\begin{eqnarray}
	\Invariant &\sdef&
	\distinctq \land
  	\injective{\reserved} \land \auxhead \notin \ran \reserved \land
	\notag
	\\ &&
	\getstatus{\auxhead} = \Granted \land
	(\forall \tidi \in q \spot \getstatus{\reservedi} = \Pending) \land
	\notag
	\\&&
	\auxhead \cons \fmap{\reserved}{\queue} =
		\fmap{\prev}{\queue} \cat [\tail] 
		\labeleqn{aux-reln}
	\\
	\notag
	\contract{\tidi } &\sdef&
	\ID{
	\curnodei
	,
	\nextnodei
	,
	\reservedi 
	,
	\getstatus{\reservedi} 
	}
	\land 
	\\ &&
	\labeleqn{advance-in-queue}
	(\tidi  \in \queue \iff \tidi  \in \queue') \land
	(\self \in \queue \imp \indexin{\queue'}{\tidi} \leq \indexin{\queue}{\tidi})
	\\
	\notag
	\rely{_i} &:&
	\Invariant \land \contract{i} 
	\\
	\notag
	\guar{}{_i} &:&
	\Invariant \land (\forall j \neq i \spot \contract{j}) 
	\\
	\notag
	\cinv &:&
	\lock = (\If \queue = \elist \Then \free \Else \hdq)
\end{eqnarray}

\caption{seL4 CLH proof structure}
\labelfig{CLH-seL4-proof}
\end{figure}

\paragraph{Acquire.}
The atomic swap instruction is
represented as an atomic sequence of instructions enclosed in angled brackets, $\atomic{\ldots}$.
\begin{equation*}
   \prev \asgn \swap{\addr\tail}{\curnodei} 
   \leadsto
   \atomic{\prev \asgn \tail ; \tail \asgn \curnodei}
\end{equation*}
The $\prev$ local is updated to the current value of $\tail$, which is then updated to thread $\tidi$'s node.
Additionally, in \reffig{CLH-seL4-proof}, the abstract queue $\queue$ is updated with $\tidi$ appended to the end, indicating the concept
of $\tidi$ joining the queue of waiting threads.
The rest of the code corresponds to that of \reffig{CLH-seL4-code}.

\paragraph{Release.}
The relatively benign-seeming line of code to set the status of the node of the releasing thread to $\Granted$
in fact holds much of the subtlety of the data structure.
This is reflected in the abstract swap of the $\reserved$ node for thread $\tidi$ with what is currently 
the abstract $\auxhead$ variable.
Thus, thread $\tidi$ takes the now-unused former head of the node queue for itself, to be used in its next attempt to acquire the lock;
and it simultaneously leaves behind its own node to become the current head of the node queue.
Abstractly, this is reflected in the queue of threads by removing the first node -- which is $\tidi$ -- from $\queue$.
The $\reserved$ and $\auxhead$ auxiliary variables are only modified at this key line of $\releaseproc$,
which straightforwardly ensures that $\reserved$ remains injective, and that $\auxhead$ is not reserved.

\paragraph{Data type invariant.}
The $\Invariant$ describes the relationships of the different variables in the system;
in particular, it outlines the 
auxiliaries $\auxhead$ and $\reserved$,
properties of $\queue$,
and the $\status$ of nodes.
The key aspect in this regard is that $\reserved$ nodes of threads currently in the queue
are always $\Pending$, while the $\auxhead$ node is always $\Granted$.
Additionally, the $\reserved$ mapping is always injective, and $\auxhead$ is never reserved by a thread.
Finally,
as shown in \refeqn{aux-reln},
the relationship between the abstract queue $\queue$ and the concrete state is that
the list of nodes obtained by reading from the auxiliary head through the reserved nodes
in the queue, 
$\auxhead \cons \fmap{\reserved}{\queue}$,
is equal to taking the $\prev$ node of each thread in the queue, with $\tail$ appended to the end,
$\fmap{\prev}{\queue} \cat [\tail]$.
This relationship gives a direct, list-based approach to describing CLH's pseudo-linked list,
and could form
the ``coupling invariant'' in a data-refinement style proof
\cite{HeHoareSanders-DR,Morgan-Gardiner:90,deRoever-DR}.

\paragraph{Contract between threads.}
The $\contracti$ between threads describes \emph{changes} that may be made to the state.
It is thus a relation, or two-state predicate, in contrast to the single-state $\Invariant$.
The contract amounts to not changing the variables assigned to a particular thread $\tidi$
(we use $\id(x,y,\ldots)$ for the relation that does not alter $x$, $y$, \etc);
and additionally, that
the $\status$ of the $\reserved$ node is never changed
(establishing this key fact without an explicit $\reserved$ variable proved cumbersome).
The contract also contains those parts already described in the specification of a queued lock from \reffig{queued-lock-spec}:
a thread cannot be entered into or removed from the queue by another thread,
and a waiting thread only ever gets closer to the front of the queue.

\paragraph{Properties of an abstract lock.}
The status of the system lock may be derived from the auxiliary queue as in \reffig{queued-lock-spec}:
if the queue is empty no thread holds the lock, and otherwise the thread at the head of the queue holds the lock
($\cinv$).

\paragraph{Relies and guarantees.}
Each thread $\tidi$ relies on its contract being fulfilled, and that the invariant is maintained;
and conversely, guarantees to fulfil the contract for every other thread in the system and to maintain the invariant.
Establishing that the guarantee is maintained by each line of code is the most complex part of the proof;
establishing the postconditions is relatively straightforward by careful choice of intermediate states.

\paragraph{Pre/postcondition and intermediate states for $\acquireproc$.}
The precondition for $\acquirei$ is that $\tidi$ is not already in the queue,
and that the concrete $\curnode$ entry accurately holds $\tidi$'s unique reserved node.
The intermediate states in \reffig{CLH-seL4-proof}
are relatively straightforward updates of this initial precondition based on the relevant assignment.
At each point we track the $\reserved$ node for $\tidi$, which is kept in $\curnodei$.
Implicitly, each intermediate state preserves the $\Invariant$.
The non-trivial part is establishing that the current thread is indeed the head of the abstract queue
when the loop terminates; this is derivable
from the $\Invariant$, specifically \refeqn{aux-reln} and the status of waiting nodes.
\begin{eqnarray}
	\tidi \in \queue \imp (\getstatus{\prev_\tidi} = \Granted \iff \tidi = \hdq)
	\labeleqn{status-prev-i}
\end{eqnarray}

\paragraph{Pre/postcondition and intermediate states for $\releaseproc$.}
The $\releasei$ operation assumes that $\tidi$ is at the head of the abstract queue,
that $\curnodei$ holds the reserved node, and that $\nextnodei$ points to the auxiliary head of the queue.
After the first line of $\releasei$, thread $\tidi$ is no longer in the queue, and has taken the former
head of the queue as its reserved node; and this node may be found in $\nextnodei$ 
(one cannot access $\previ$
directly since it is local to $\acquirei$).
The thread therefore updates $\curnodei$ to $\nextnodei$,
ready for the next call to $\acquirei$.
As with $\acquirei$, at each state the $\Invariant$ is implicitly maintained.

\subsection{Thread and system proofs}

\newcommand{\spacedquintI}[6]{\quintI{#1}{#2}{#3}{#4}{\\ \phantom{\{} #5}{#6}}
\newcommand{\spacedquintIx}[6]{\quintI{#1}{\\ \phantom{\{} #2}{#3}{#4}{#5}{#6}}

We now state the top-level theorems showing that the \clhsel implementation is correct.
All proofs have been encoded and machine-checked using Isabelle/HOL.
We include an invariant in the quintuple statements \refeqn{invariant-quintuple},
which adds the invariant to pre/postconditions and rely/guarantee relations.
\begin{theorem}[Acquire]
\labelth{acquire}
\[
\hskip -4mm
	\quad
	\spacedquintI{
   		\self \notin \queue \land \curnodei = \reservedi}
		{\contracti}
	{
	\\
\hskip -4mm
   		\acquirei
	}{\forall \tidj \neq \tidi \spot \contractj}{
		\self \in \queue \land \curnodei = \reservedi \land \nextnodei = \auxhead 
		 \land \self = \hdq
	}{
		\Invariant
	\\ 
\hskip -4mm
\quad
	}
\]
\end{theorem}
\begin{theorem}[Release]
\labelth{release}
\[
\hskip -4mm
	\quad
	\spacedquintIx{
		\self \in \queue \land \curnodei = \reservedi \land \nextnodei = \auxhead \land \self = \hdq
	}
		{ \contracti}
	{
	\\
\hskip -4mm
   		\releasei
	}{\forall \tidj \neq \tidi \spot \contractj}{
   		\self \notin \queue \land \curnodei = \reservedi
	}{
		\Invariant
	\\ 
\hskip -4mm
\quad
	}
\]
\end{theorem}

\begin{clhproof}
Individual lines of code
are shown to satisfy their pre/post states
outlined in \reffig{CLH-seL4-proof}
using the inference rules for assignments (\refinfrule{asgn}), and spin loops (\refinfrule{spin-loop}), 
and are joined using \refinfrule{seq} using the relevant mid-states.
The relies and guarantees are kept constant throughout.
\end{clhproof}

The top-level property for 
a system of $\numThreads$ threads calling $\acquireproc$ and $\releaseproc$ repeatedly is given below.
The initial state is derived from that given at the top of \reffig{CLH-seL4-proof}.
The overall postcondition (we leave unconstrained the condition under which a thread may finish execution)
shows that the queue is empty, though the rest of the state could have arbitrary values.
Note that the guarantee $\forall \tidi \spot \contracti$
is a simplification of $\forall \tidi \spot \forall \tidj \neq \tidi \spot \contract{\tidj}$.
These simplifications of pre/postconditions and rely/guarantees follow from \refinfrule{conseq}.
\begin{theorem}[Locking system]
\labelth{toplevel-system}
\[
	\quad
	\quintI{
		\queue = \elist \land \reserved = \ldots
	}{
		\forall \tidi \spot \contracti
	}{
	\\
		\genpar{\tidi}{\While \_ \Do (\acquirei \Seq \releasei)}
	}{
		\forall \tidi \spot \contracti
	}{
		\queue = \elist \land \ldots
	}{
		\Invariant
	\\ \quad
	}
	\\
\]
\end{theorem}
\begin{clhproof}
We apply \refinfrule{par-gen}:
each individual thread satisfies its own part from \refths{acquire}{release}
(and that the postcondition of $\acquirei$ is the precondition of $\releasei$, using \refinfrule{seq});
careful choice of $\rely_i$ and $\guar_i$ from \reffig{CLH-seL4-proof}
ensure that each thread's guarantee fulfils the rely of every other thread.
\end{clhproof}

For simplicity,
\refth{toplevel-system}
explicitly uses the contract between individual threads; however, from the perspective 
of a system that incorporates the lock,
the rely of the lock system is simply that none of its variables are modified, 
and conversely the lock system guarantees to change nothing other than its own variables.
These are both implied by $\forall \tidi \spot \contracti$.

That the lock itself is used in such a way as to guarantee mutual exclusion to some system variable $x$
is something that must be shown for each client system, 
\ie that $x$ is only ever modified between calls to $\acquirei$ and $\releasei$.



\newcommand{\acquiresel}{\acquireproc^{\normalfont\textsc{sel4}}}
\newcommand{\acquireseli}{\acquiresel_\tidi}
\newcommand{\acquiremod}{\mathsf{acquire}^{\pl}}
\newcommand{\acquiremodi}{\acquiremod_\tidi}

\newcommand{\releasesel}{\releaseproc^{\normalfont\textsc{sel4}}}
\newcommand{\releaseseli}{\releasesel_\tidi}
\newcommand{\releasemod}{\mathsf{release}^{\pl}}
\newcommand{\releasemodi}{\releasemod_\tidi}

\newcommand{\wh}{\kw{wait}}
\newcommand{\tmpvar}{\mathtt{t}}

\section{CLH-seL4 with Weak Memory}
\labelsect{wmms}
\newcommand{\armfence}{\mathtt{dsb}}

\renewcommand{\ffence}{\kw{fence}}

The CLH lock is an algorithm that may be deployed at the lowest level of the software stack,
and thus has little abstraction from hardware-level concerns.
Chief among these hardware-level factors is \emph{out-of-order execution}, which has been a feature of processors since the 1960s.
Processors may execute machine instructions out-of-order to
make the best use of resources. For instance, given consecutive, 
\emph{independent} instructions -- \eg arithmetic operations accessing different sets of registers --
it is more efficient to execute them in parallel than wait for the first to complete before the second begins --
\eg if multiple arithmetic units are available.

The key principle of out-of-order execution is that accesses to the same variable $x$ must be kept in-order, while ``independent'' instructions (\eg accessing different parts of memory) may be reordered.
On a single processor, this principle maintains the correctness of sequential code while improving performance
by eliminating unused processor cycles.
However, the effects of parallelisation can
become apparent on modern multicore architectures, with potentially disastrous effects on correctness.

\subsection{Identifying Weak Memory Effects}

We analyse the possible weak memory effects in a piece of code by searching for instructions that may be executed out of order.
For such instructions, we either
(1)
transform the code to make this parallelism explicit, where it does not affect correctness, or 
(2) 
insert fences or other ordering constraints to enforce the required ordering.

For a given memory model $\mm$ (for instance, Arm, x86, RISC-V, or \Clang),
following \cite{colvin-SEFM21},
we write $\aca \rom \acb$ if instruction $\acb$ may be reordered with instruction $\aca$ under memory model $\mm$.
Conversely, we write
$\aca \nrom \acb$ if they may not be reordered, which may be because they access the same variables,
or if either instruction includes some sort of artificial ordering constraint, such as a fence.

\newcommand{\transM}[2]{#1^{+#2}}
\newcommand{\cm}{\transM{c}{\mm}}

If a sequential program $c$ is subject to a memory model $\mm$, 
its set of traces is determined by the (weak memory) semantics given in \cite{colvin-SEFM21} after 
transforming $c$ by interpreting each sequential composition ``;'' as a \emph{parallelized sequential composition}
``$\ppseqm$'', which makes the inclusion of the memory model $\mm$ explicit.

Following
\cite{colvin-SEFM21}
we transform code that uses parallelized sequential composition according to refinement rules based on trace equivalence.
The two key rules are the following:
\begin{eqnarray}
	\aca \nrom \acb 
	~~\imp~~
	\aca \ppseqm \acb \aeq \aca \Seq \acb
	\labeleqn{nrom}
	\\
	\aca \rom \acb 
	~~\imp~~
	\aca \ppseqm \acb \aeq \aca \pl \acb
	\labeleqn{rom}
\end{eqnarray}
If two instructions cannot be reordered according to $\mm$, then 
they are executed in order, exactly as if familiar sequential composition was used \refeqn{nrom}.
Conversely, if two instructions may be executed out of order according to $\mm$,
then effectively they are executed in parallel \refeqn{rom}.%
\footnote{\emph{Forwarding} needs to be taken in account in general, but does not occur in CLH.}
Note that this second rule
may result in nested parallelism within a process.

Most weak memory models include some sort of barrier or fence instruction ``$\ffence$'' that
can be used to reinstate sequential order.  
This is summarised by the following rule.
\begin{equation}
	\labeleqn{ffence}
	c_1 \ppseqm \ffence \ppseqm c_2
	~~=~~
	c_1 \Seq c_2
\end{equation}
Note that we may eliminate both the fence and the top-level references to $\mm$, reducing parallelized sequential composition
to standard sequential composition.
Similar rules apply for other ordering constraints, such as \Clang's \T{memory\_order} type.
More complex rules can be derived for conditionals and loops \cite{colvin-SEFM21,colvin-ICFEM22}.

\subsection{Weak Memory Effects in Acquire}
\labelsect{wmm-acquire}

\renewcommand{\r}{\concrete{r}}
\newcommand{\procsi}{\procs[\tidi]}

\newcommand{\ia}{\mathtt{\r \asgn \procsi.\curnode } }
\newcommand{\ib}{\mathtt{\getstatus{\r} := \Pending }}
\newcommand{\ic}{\mathtt{\prev \asgn \swap{\addr\tail}{\r}}}
\newcommand{\itmp}{\mathtt{\procsi.\nextnode \asgn \prev}}
\newcommand{\iwh}{\mathtt{\awaitgranted}}

{

To address weak memory effects on CLH we consider the interactions between different lines of code.
At this level, we must consider the atomicity of instructions explicitly. 
To more realistically represent the code we make the loading of $\curnodei$ into a local
variable (register) $\r$ explicit, \ie, we introduce a new (atomic) action.
The individual lines of $\acquireproc$ from \reffig{CLH-seL4-code} are repeated below with 
potential reorderings between consecutive lines made explicit.
\begin{align}
		&
	\ia
		& \nroA
	\\
		&
	\ib
		& \roA
	\\
		&
   	\ic
		& \nroA
	\\
		&
	\itmp
		& \roA
	\\
		&
	\iwh
\end{align}
The first two instructions may not be reordered: the first modifies $\r$, and the second
reads $\r$.  Note that the expression $\getstatus{\r}$ implicitly dereferences $\r$,
requiring an update of the location pointed to by $\r$.  
The $\swapproc$ operation 
(which we interpret according to the atomic statement in \reffig{CLH-seL4-proof})
modifies both the local $\prev$ variable and the global $\tail$ variable, and also reads 
(but does not modify) $\r$.
As above, the instruction
$
	\ib
$
reads but does not modify $\r$.  Since $\r$ is local,
these instructions can reorder according to the natural constraints of the Arm memory model
(and indeed, almost any other).
Treating pointer dereferences as a load (of a variable) and a store (to a location) is
covered by the framework in \cite{colvin-FM23}.

The $\swapproc$ operation cannot reorder with the initial update of $\r$,
that is,
$\ia \nroA \ic$
(otherwise the sequential semantics would be compromised).
Hence the first three instructions, executing under Arm,
is trace equivalent to
the following, by \refeqns{nrom}{rom}.
\[
	\ia \Seq (\ib \pl \ic)
\]
If the $\swapproc$ occurs first, this could lead to a violation of the part of the invariant
requiring that nodes in the queue have status $\Pending$, and invalidate the key property
\refeqn{status-prev-i}.
We must therefore enforce ordering between the setting of $\status$ and the $\swapproc$. 
For Arm (or \Clang), ordering in this case can be enforced by inserting a fence (such as $\armfence$),
or by placing a ``\emph{release}'' ($\release$) ordering constraint on the latter of the two stores
\cite{ReleaseConsistency90}.

The instruction immediately following the $\swapproc$, $\itmp$,
accesses local $\prev$, and because $\prev$ is modified by the $\swapproc$
those two instructions remain strictly ordered.

The succeeding $\Await$ loop can be interpreted as a sequence of loads of the global location $\getstatus{\prev}$.
All such loads are independent of the previous store to $\procsi.\nextnode$
(but, since the store and the loop both access local $\prev$, they are dependent on the $\swapproc$).
As such, the loop and the preceding assignment may effectively happen in parallel
(including if a compiler decides to restructure the assignment 
after the loop).
\begin{equation}
	\itmp \pl \iwh
	\labeleqn{pl-while}
\end{equation}
This makes no difference to the correct functioning of the code and there is no need to enforce this ordering,
in contrast to the potential reordering involving the $\swapproc$ above.

\begin{theorem}[Parallelization of the while loop in acquire]
\labelth{par-sel4}
\\
$
	~~~
	\quintI{
   		\self \in \queue \land \curnodei = \reserved[\self]
   	}
	{\contracti}
	{
	\\
	\itmp \pl \iwh
	}{
	\forall \tidj \neq \tidi \spot \contract{\tidj}
	}{
		\\ \quad \phantom{\{} \quad
   		\self \in \queue \land \curnodei = \reserved[\self] \land \nextnodei = \auxhead \land \self = \hdq
	}{
	\Invariant
	\\ ~~~
	}
$
\end{theorem}
\begin{clhproof}
For such proofs involving nested parallelism we apply \refinfrule{par-u} (as opposed to \refinfrule{par-int}).
In this case the left-hand side of the parallel composition, 
$\itmp$
inflicts more interference on the while loop than we had taken into account in \reffig{CLH-seL4-proof}.
However, \refinfrule{spin-loop} still applies -- the extra interference by the potential change to $\procsi.\nextnode$
does not compromise the stability of the pre or post condition.
Furthermore, being a spin loop, the right-hand side of the parallel composition does not inflict any more interference
on the left-hand side assignment than had previously been accounted for.
\end{clhproof}

The pre and post conditions in \refth{par-sel4}
match the pre and post conditions for the corresponding sequential composition in \reffig{CLH-seL4-proof} --
only the midstate is missing.
We can therefore replace the sequential composition with parallel composition and maintain top-level correctness.
Whether the assignment occurs before or after the loop makes no difference,
which is precisely the reason why such reorderings are allowed by processors and compilers.

We may therefore determine that the only necessary artificial ordering constraint that must be
injected into the original sequential code for \clhsel 
is to enforce ordering between the $\Pending$ status assignment and the $\swapproc$.
All other possible reorderings allowed by the memory model are irrelevant to correctness.  
We summarise this below, where for space
we abbreviate
$
	\wh
	\sdef 
	\iwh
$.
\begin{eqnarray}
\labeleqn{acquiresel}
	\acquireseli
	\asdef
	\r \asgn \procsi.\curnode
	\ppseqA
	\getstatus{\r} \asgn \Pending
	\ppseqA
	\\ && 
	\notag
	\prev \asgn \swap{\addr\tail}{\r}^{\release}
	\ppseqA
	\procsi.\nextnode \asgn \prev
	\ppseqA
	\wh
	\\
	\acquiremodi
	\asdef
	\r \asgn \procsi.\curnode
	\Seq
	\getstatus{\r} \asgn \Pending
	\Seq
	\\ && 
	\notag
	\prev \asgn \swap{\addr\tail}{\r}
	\Seq
	(
	\procsi.\nextnode \asgn \prev
	\pl
	\wh
	)
\end{eqnarray}

\begin{theorem}[Acquire under $\ARMmm$]
\labelth{acquire-arm}
By trace equivalence
modulo fences and $\release$ ordering constraints,
$
	\acquireseli
	=
	\acquiremodi
$.
\end{theorem}
\begin{clhproof}
As described above, the $\release$ constraint on the swap operation is sufficient to enforce sequential ordering 
and may be replaced by strict sequential composition,
and in the absence of any other ordering constraints the assignment before the loop and the loop itself may be parallelized.
We have therefore reduced all the instances of $\ppseqA$ to either sequential or parallel composition, in line with that given in $\acquiremodi$.
\end{clhproof}

\subsection{Weak Memory Effects in Release}
\labelsect{wmm-release}

\renewcommand{\p}{\procs[\tidi]}

The two lines of code in $\releaseproc$ involve the loading of $\curnodei$ followed by the modification of the same; hence
the key parts of the two instructions are kept strictly ordered by any memory model
(breaking each of the two lines of code in $\releaseproc$ into their atomic assembler instructions introduces some
parallelism as in $\acquireproc$, but not for the key store and load to $\curnodei$).
\begin{equation}
\labeleqn{ro-release}
 \mathtt{
    \p.\curnode\fun\status \asgn \Granted
	} 
	\nroA
	\mathtt{
	\p.\curnode \asgn \p.\nextnode
	}
\end{equation}
As such, the sequential composition in $\releaseproc$ may be treated as strict.
\begin{theorem}[Release under $\ARMmm$]
\labelth{release-arm}
$
	\releaseseli
	=
	\releasei
$.
\qed
\end{theorem}
\begin{clhproof}
Immediate from \refeqn{ro-release} and \refeqn{nrom}.
\end{clhproof}

\subsection{Reordering with the critical section}

The reasoning above is sufficient for considering the code of each operation in isolation; 
however, further reorderings with code between the calls to $\acquireproc$ and $\releaseproc$
are possible.  Essentially \emph{any} reading or writing to main memory before $\acquireproc$
has fully completed (been flushed from the pipeline) or while $\releaseproc$ is being executed is
problematic from the perspective of maintaining mutual exclusion in the critical section,
as outlined in \refsect{lock-props}.
For instance, in a calling context, the code $\acquirei \ppseqA critical~section \ppseqA \releasei$ could allow
parallelisation between the instructions of $\acquirei$ and shared-variable accesses in the critical section,
and even with instructions of $\releasei$.
If the implementation of $\acquireproc$ includes a full fence (in Arm, the \armdsb instruction would suffice)
as the final instruction, 
and similarly the implementation of $\releaseproc$ includes a fence as its first instruction,
by \refeqn{ffence}
we may apply
standard sequential reasoning.
\begin{eqnarray}
	\acquirei \ppseqA crit \ppseqA \releasei 
	~~\aeq ~~
	\acquirei \Seq crit \Seq \releasei
\end{eqnarray}

In contrast,
code prior to a call to $\acquireproc$ need not be fenced, under the reasonable assumption that
no code other than that of $\acquireproc$ and $\releaseproc$ modifies the variables in \reffig{CLH-seL4-code}.
Similar reasoning holds for code after a call to $\releaseproc$;
in both cases, under reasonable assumptions about access to the lock-specific variables, no further barriers to reordering are necessary.

A possible exception is if thread $\tidi$ makes a call to $\releaseproc$ immediately prior to a call to $\acquireproc$.
Again, a simple analysis of the first instruction of $\acquireproc$ and the last of $\releaseproc$
shows that both access $\curnodei$ and hence reordering is not possible.  Thus we may establish
$\releasei \ppseqA \acquirei = \releasei \Seq \acquirei$.

}

\subsection{Top-level proof}

\begin{theorem}[Correctness of CLH under $\ARMmm$]
The implementation of CLH in \reffig{CLH-seL4-code}, updated with a $\release$ constraint as in \refeqn{acquiresel}, and augmented 
with $\armfence$ barriers (or equivalent) at the end of $\acquireproc$ and at the beginning of $\releaseproc$,
works as a lock under $\ARMmm$.
\end{theorem}
\begin{clhproof}
By \refth{acquire}, \refth{par-sel4}, and \refth{acquire-arm},
the implementation of $\acquireproc$ satisfies the pre and postconditions of \reffig{CLH-seL4-proof};
and similarly for the implementation of $\releaseproc$ by \refth{release} and \refth{release-arm}.
These may be composed identically to \refth{toplevel-system}.
The definition and validity of the interpretation of the Arm memory model is taken from \cite{colvin-SEFM21}.
\end{clhproof}

Taking into account weak memory models (in this case, Arm, though the reasoning applies to essentially any real model, such as RISC-V and x86-TSO
\cite{colvin-SEFM21}),
we find that artificial orderings must be inserted to enforce the order of the initialisation of the status of the local node to $\Pending$
and the atomic swap; however, the standard constraints of standard memory models either enforce ordering of statements, 
or, in the case of the loop, the stated program order of the statements is not strictly necessary for correctness.
The required additional reasoning is separate to the application of rely/guarantee reasoning.



\section{Related work}
\labelsect{related-work}

The verification of concurrent algorithms, including locks
\cite{Hesselink-PVS-Bakery,MCS-CertiKOS,Schellhorn-2016},
has a long history comprising a variety of techniques and tools.
Approaches that have been applied specifically to
CLH include several works which have focussed on termination/liveness properties using separation logics \cite{TaDA-Live,LiLi,CLH-ghost}.
The application of the Starling framework to CLH \cite{CLH-starling} focusses on functional correctness, with an underlying technique based on Owicki-Gries reasoning
\cite{OwickiGries76}.  
It is unclear how these works can be adapted to handle the extra complexity associated with weak memory concerns, which must be taken into account when 
considering deployment of CLH on modern devices.
As we show in \refsect{wmms}, some but not all of the parallelism that multicore architectures introduce can be retained.  A verification approach
that instead insists on restoring full sequential ordering would therefore introduce unnecessary inefficiencies associated with the overuse of fences.
We have not addressed termination directly in this paper, however we can use the results of \refsect{wmms} to enable
the application of standard techniques; we have made a 
step towards that by showing that every process is guaranteed to only ever advance in the abstract queue \refeqn{advance-in-queue}.
By basing our approach on rely/guarantee reasoning we get compositionality -- helpful for managing micro-parallelism of weak memory effects --
and an existing machine-checked theory base in Isabelle/HOL \cite{RGinIsabelle}.

If we were instead to attempt to verify CLH directly using a weak-memory-specific technique
such as
\cite{Views-inproc,Dongol-Unifying,PromisingSemanticsKang}
we would typically need to rework the verification conditions (pre/postconditions) into a technique-specific assertion language,
designed to capture weak memory effects, and apply technique-specific inference rules.
In our approach, weak memory effects are elucidated separately to the main verification process.
Furthermore, we may tackle other properties of CLH --
for instance, liveness and security, 
both of importance in the context of applications such as seL4
--
by applying standard techniques to the transformed code;
otherwise, new assertion languages and inference systems need to be devised to handle these additional concerns under the effects of weak memory models.
The approach of 
\cite{Coughlin-Smith-2022}
is based on a similar semantic model to ours, but is, again, specific to a particular verification technique;
however, it has been adapted to handle the Power architecture, which includes other micro-architectural features that complicate correctness
\cite{Coughlin-Winter-Smith-2023}.
Our separation-of-concerns approach has a further advantage in that the inclusion of auxiliary variables 
does not complicate the reasoning about weak memory effects.



\section{Conclusions}
\labelsect{conclusions}

We have shown an end-to-end machine-checked verification of a complex, practical, in-use concurrent algorithm for multicore architectures
such as Arm.
The approach is centred on Jones' rely/guarantee concept, which is suitable
for describing top-level specifications (\refsect{lock-specification}),
detailed proofs of individual threads (\refsect{CLH-seL4-proof}),
and compositional reasoning about inter- and intra-thread parallelism (\refsect{wmms}).
We utilise and specialise mature tool support for theorem proving using rely/guarantee concepts
\cite{RGinIsabelle,IsabelleHOL,Paulson:94}
(theory files available on request).

In constructing the proof in \refsect{CLH-seL4-proof} we also proved several simpler algorithms, including:
abstract versions of locks, one with only a single $\lock$ variable, and one with an explicit $\queue$ variable
as in \refsect{lock-specification};
two versions of an array-based
queue lock (which suffers from space considerations when multiple locks are used within a system); 
and an alternative version of CLH given by Scott \cite{scott-13}. 
In this last proof we were able to reuse many of the results from \refsect{CLH-seL4-proof},
and indeed to show that both implementations are \emph{compatible}, in the sense that different threads could 
use the different implementations without clashing.  This reuse is owed in large part to abstracting the details in
auxiliary variables and the inherent compositionality of rely/guarantee reasoning, as well as separating the weak memory
effect reasoning (again, much of which was reused).
A more typical approach is to use program counters
or
the application of Owicki-Gries,
both of which would make proof reuse across different implementations much harder.

We are currently working towards the automation of the program-level reasoning in \refsect{wmms}
and tool support for \emph{derivational} rely/guarantee proofs \cite{SRA16,SemanticsSRA}.


{\bf Acknowledgements.}
This project was funded by the Department of Defence and administered through the Advanced Strategic Capabilities Accelerator.
We thank David Gwynne for help with understanding the efficiency implications of CLH.

\bibliographystyle{splncs04}
\bibliography{biblio.bib}

\end{document}